\documentclass[10pt]{article}
\usepackage{latexsym}
\usepackage{amssymb}
\usepackage{amsfonts}
\usepackage{amsmath}
\usepackage{theorem}

\newtheorem{theorem}{Theorem}
\newtheorem{lemma}{Lemma}

\newtheorem{definition}{Definition}

\begin{document}

\title{Random Multi-Overlap Structures and\\ 
Cavity Fields in   
Diluted Spin Glasses}
\author{Luca De Sanctis
\footnote{Department of Mathematics,
Princeton University, Fine Hall, Washington Road,
Princeton NJ 08544--1000 \ USA \ 
{\tt<lde@math.princeton.edu>}}}

\maketitle

\begin{abstract}
We introduce the concept of Random Multi-Overlap Structure
(RaMOSt) as a generalization of the 
one introduced by M. Aizenman
R. Sims and S. L. Starr for non-diluted spin glasses. We use 
such method
to find generalized bounds for the 
free energy of the Viana-Bray model of
diluted spin glasses and to
formulate and prove the 
Extended Variational Principle
that implicitly provides the free energy 
of the model. Then we exhibit
a theorem for the limiting RaMOSt,
analogous to the one found by F. Guerra for the
Sherrington-Kirkpatrick model, that describes some
stability properties of the model. Last,
we show how our technique can be used 
to prove the existence of thermodynamic 
limit of the free energy. The present 
work paves the way to a revisited
Parisi theory for diluted spin systems.
\end{abstract}



\section{Introdution}

The diluted mean field spin glasses are 
important both for their correspondence
to random optimization problems and for 
their sort of intermediate nature halfway 
from idealized mean field models to 
short range realistic ones, thanks to the 
finite degree of connectivity.

Among the few rigorous results obtained 
so far in diluted spin glasses, two
important examples are ref. \cite{franz, t1},
where S. Franz 
and M. Leone found bounds 
for the free energy of 
diluted spin systems, considering the first level
of Replica Symmetry Breaking;
while D. Panchenko and M. Talagrand
found a way to consider any level 
Broken Replica Symmetry Bound in a compact way,
using a weighting scheme inspired by ref. \cite{ass}.
In the high temperature region, 
rigorous results have been
obtained for the K-Sat model of diluted 
spin glass by M. Talagrand$^{\cite{t2}}$ and 
by F. Guerra and F. L. Toninelli
for the Viana-Bray model$^{\cite{gt2}}$.

In the case of non-diluted models,   
M. Aizenman, R. Sims and S. L. Starr  
recenlty 
introduced the concept of
ROSt (Random Overlap Structures) 
through which they found bounds
for the free energy in a very
elegant and easy manner. 
In the same important paper$^{\cite{ass}}$, 
the authors expressed 
the solution through an 
Extended Variational Principle. 
An important restriction of the 
ROSt space  
has been done by F. Guerra$^{\cite{g2}}$, 
exhibiting invariance of the limiting 
ROSt under certain transformations. 

After the introduction of the basic definitions in
section \ref{notations}, we extend the ideas of ref. \cite{ass}
to diluted spin glasses, in section \ref{bounds}.
The finite connectivity
requires that we
consider Multi-Overlap as opposed to
Overlap Structures (because the couplings
are not Gaussian). In section \ref{vp} we prove a generalized bound 
for the free energy of the 
Viana-Bray model, by means of an
interpolation not 
based on the iterative approach of ref. \cite{parisi2} 
used to find bounds 
in ref. \cite{franz, t1}. Rather, our interpolation is closer to the 
one used for non-diluted models in ref. \cite{ass}. 
As a consequence we can (like in ref. \cite{ass})
formulate an Extended 
Variational Principle for the free energy. The
next natural step we performed, in section \ref{optimal}, is the search
for invariant transformations of the optimal limiting RaMOSt, 
and we found stability properties similar to those found 
for non-diluted systems in ref. \cite{g2}. Appendix \ref{a} is devoted
to a calculation that plays a basic role throughout the paper,
Appendix \ref{tl} contains a somewhat new proof of the existence of the thermodynamical 
limit of the free energy, Appendix \ref{srtf} reports some comments
about optimal versus non-optimal RaMOSt's in terms of the phenomenon of
overlap coalescence and generalized trial functions.


\section{Model, Notations, Definitions}\label{notations}

We refer to ref. \cite{gt2} for an introduction to 
the Viana-Bray model, 
a physical description,
the role of replicas and multi-overlaps, the infinite 
connectivity limit and the connection 
to the Sherrington-Kirkpatrick model, 
the behavior in the annealed region.\\
We will have in mind a lattice with a large bulk of $N$ sites 
(cavity) and $M$ additional spins
($N$ is large and $M$ is fixed).

Notations: \\
$\alpha, \beta, h$ are 
non-negative real numbers
(degree of connectivity, inverse temperature and 
external field
respectively);\\
$P_\zeta$ is a Poisson random variable of mean $\zeta$; \\
$\{i_\nu\}, \{l_\nu\}$ are independent identically 
distributed random variables, 
uniformly distributed over the cavity points $\{1,\ldots, N\}$;\\
$\{j_\nu\}, \{k_\nu\}$ are independent identically 
distributed random variables, uniformly 
ditributed over the added points $\{1,\ldots, M\}$;\\
$\{J_\nu\}, \{\hat{J}_\nu\}, \{\tilde{J}_\nu\}, J$ are 
independent identically distributed 
random variables, with symmetric distribution;\\
$\mathcal{J}$ is the set of all the quenched 
random variables  above;\\
$\sigma: j \rightarrow \sigma_{j}=\pm 1, \tau: i \rightarrow \tau_i=\pm 1$ 
are the added 
and cavity spin configurations respectively; 
$\rho_.$ will be used 
for the spins in the full lattice (the points of which
are denoted by $r , s$) 
without 
distinguishing between cavity and added spins;\\
$\pi_{\zeta}(\cdot)$ is the Poisson measure
of mean $\zeta$.\\ 
$\mathbb{E}$ is an average over all (or some of) the 
quenched variables;\\
$\omega_\mathcal{J}$ is the Bolztmann-Gibbs 
average explicitly written below;\\
$\Omega_K$ is a product of the needed 
number of independent identical copies (replicas) 
of $\omega_\mathcal{J}$,
in a system with $K$ spins;\\
$\langle\cdot\rangle$ will indicate the composition 
of an $\mathbb{E}$-type average over some 
quenched variables and some sort of 
Boltzmann-Gibbs average over the spin variables, 
to be specified each time.\\
We will often drop the dependance on 
some variables or indices or slightly
change notations to lighten 
the expressions, 
when there is no ambiguity.\\
In absence of external field, the Hamiltonian 
of the system of $M$ sites
is, by definition
\begin{equation*}
\label{ham}
H_M(\sigma, \alpha; \mathcal{J})=
-\sum_{\nu=1}^{P_{\alpha M}} J_\nu \sigma_{j_\nu}\sigma_{k_\nu}
\end{equation*}
When there is an external field $h$, the Hamiltonian is
$H_M+H^{ext}_{M}$, where we used the definition
$H^{ext}_{M}(\sigma, h)=-h\sum_{j=1}^{M}\sigma_{j}$.\\
We follow the usual basic definitions and notations 
of thermodynamics for the partition function and 
the free energy per site 
\begin{eqnarray*}
\label{z}
&&Z_M(\beta, \alpha, h; \mathcal{J})=\sum_{\{\sigma\}}
\exp(-\beta(H_M(\sigma, \alpha; \mathcal{J})
+H^{ext}_{M}(\sigma, h))),\\
\label{f}
&&-\beta f_M(\beta, \alpha, h)=\frac1M \mathbb{E}
\ln Z_M(\beta, \alpha; \mathcal{J})
\end{eqnarray*}
and $f=\lim_M f_M$.\\
The Boltzmann-Gibbs average of an observable $\mathcal{O}$ is
\begin{equation*}
\omega_{\mathcal{J}}(\mathcal{O})=
Z_M(\beta, \alpha, h; \mathcal{J})^{-1}
\sum_{\{\sigma\}}\mathcal{O}(\sigma)\exp(-\beta(
H_M(\sigma,\alpha;\mathcal{J})+H^{ext}_{M}(\sigma, h)))
\end{equation*}
The multi-overlaps are defined (using replicas) by
\begin{equation*}
\label{overlap}
q_{n}=\frac{1}{N}\sum_{i=1}^N\tau_i^{(1)}\cdots\tau_i^{(n)}\ ,\  
\tilde{q}_{n}=\frac{1}{M}\sum_{j=1}^{M}
\sigma_j^{(1)}\cdots\sigma_j^{(n)}
\end{equation*}
We are going to use the 
two following independent auxiliary Hamiltonians:
\begin{eqnarray}
\label{kappa}
&&\kappa(\tau, \alpha; \mathcal{J})=
-\sum_{\nu=1}^{P_{\alpha M}} 
\hat{J}_\nu \tau_{i_\nu}\tau_{l_\nu}\\
\label{eta}
&&\eta(\tau, \sigma, \alpha; \mathcal{J})=
-\sum_{\nu=1}^{P_{2\alpha M}} 
\tilde{J}_\nu \tau_{i_\nu}\sigma_{j_\nu}\equiv 
\sum_{j=1}^{M}\eta_{j}\sigma_{j}
\end{eqnarray}
where $\eta_{j}$ is the \emph{Cavity Field} 
acting on $\sigma_{j}$ defined by
\begin{equation*}
\label{ }
\eta_{j}=\sum_{\nu=1}^{P_{2\alpha}}J_{\nu}^{j}\tau_{i_{\nu}^{j}}
\end{equation*}
and the index $j$ of $J_{\nu}^{j}$ and $\tau_{i_{\nu}^{j}}$ de-numerates 
independent copies of the corresponding random variables.\\
The two expressions of $\eta$ define the same random variable, but
the first is probably the most convenient for the calculations
in next two sections,
while the second describes better the physics of the model,
and will be essential in section \ref{optimal}.


\section{Generalized Bound for the Free Energy}\label{bounds}

For $t\in [0, 1]$, consider the following 
\emph{Interpolating Hamiltionian}
\begin{equation*}
\label{ }
H(t)=H_M(t\alpha)+\kappa(t\alpha)+\eta((1-t)\alpha)+H^{ext}_{M}
\end{equation*}
and using a set of weights $\xi_{\tau}$ define 
\begin{equation}
\label{r}
R_M(t)=
\frac{1}{M}\mathbb{E}\ln
\frac{\sum_{\tau, \sigma}\xi_\tau\exp(-\beta H(t))}
{\sum_{\tau}\xi_\tau\exp(-\beta \kappa)}
\end{equation} 
Call $G_M$ the value of $R_M$ at $t=0$
\begin{equation*}
G_M(\beta, \alpha, h; \xi)=
\frac{1}{M}\mathbb{E}\ln
\frac{\sum_{\tau, \sigma}\xi_\tau\exp(-\beta(\eta+H^{ext}_{M}))}
{\sum_{\tau}\xi_\tau\exp(-\beta \kappa)}
\end{equation*}
then
\begin{eqnarray*}
\label{ }
 &&R_M(0) = G_M \\
 &&R_M(1) = -\beta f_M
\end{eqnarray*}
\begin{theorem}[Generalized Bound]
\label{b}
\begin{equation*}
\label{ }
-\beta f\leq \lim_{M\rightarrow\infty}
\inf_{\xi} G_M
\end{equation*}
\end{theorem}
\textbf{Proof}\\
The proof is based on Lemma \ref{lemma},
Appendix \ref{a}.

Define
\begin{equation}
\label{upsilon}
\Upsilon(m_{1}, m_{2}, m_{3})=
\exp(\beta (
\sum_{\nu=1}^{m_{1}}J_{\nu}\sigma_{j_{\nu}}\sigma_{k_{\nu}}
+\sum_{\nu=1}^{m_{2}}\hat{J}_{\nu}\tau_{i_{\nu}}\tau_{l_{\nu}}
+\sum_{\nu=1}^{m_{3}}\tilde{J}_{\nu}\tau_{i_{\nu}}\sigma_{j_{\nu}}))
\end{equation}
Let us compute the $t$-derivative of $R_M$, keeping in mind that
its denominator does not depend on $t$.
\begin{eqnarray*}
&&\frac{d}{dt}R_M(t)=\frac{d}{dt}\frac{1}{M}\mathbb{E}\ln
\frac{\sum_{\tau, \sigma}\xi_\tau\Upsilon(P_{t\alpha M}, P_{t\alpha M}, 
P_{(1-t)2\alpha M})}{\sum_{\tau}\xi_\tau\exp(-\beta \kappa)}=\\
&&\frac{1}{M}\sum_{\{m_{.}\}}^{0, \infty}\frac{d}{dt}
\pi_{t\alpha M}(m_{1})\pi_{t\alpha M}(m_{2})\pi_{(1-t)2\alpha M}(m_{3})
\mathbb{E}\ln\sum_{\tau, \sigma}\xi_{\tau}\Upsilon(m_{1}, m_{2}, m_{3})
\end{eqnarray*}
Now we have the sum of three terms, in each of which one of the 
$\pi$'s is differentiated with respect to $t$. As in Appendix \ref{a},
we can 
substitute into the first term  
the following relation 
$$
\Upsilon(m_{1}, m_{2}, m_{3})=
\exp(\beta J_{m_{1}}\sigma_{j_{m_{1}}}\sigma_{k_{m_{1}}}
)\Upsilon(m_{1}-1, m_{2}, m_{3})
$$
and we can do the same for the other two terms.\\
It is clear then that as in Lemma \ref{lemma} of Appendix \ref{a}
we get an average 
$\Omega_{\xi, t}$ with weights  
consisting of the weights $\xi_{\tau}$ times the 
Boltzmann-Gibbs weights associated to $H(t)$:
\begin{eqnarray*}
\hspace{-0.6cm}&&\frac{d}{dt}R_M(t)=\\
\hspace{-0.6cm}&&\alpha
[\mathbb{E}\ln\Omega_{\xi, t}
\exp(\beta J\sigma_{j_{.}}\sigma_{k_{.}})
\!+\!\mathbb{E}\ln\Omega_{\xi, t}
\exp(\beta J\tau_{i_{.}}\tau_{l_{.}})
\!-\!2\mathbb{E}\ln\Omega_{\xi, t}
\exp(\beta J\tau_{i_{.}}\sigma_{j_{.}})]
\end{eqnarray*}
According to Appendix \ref{a} we will get now some 
terms in $\cosh(\beta J)$,
such terms can be factorized out,
but here they are
cancelled since they sum up to zero.
Since
\begin{equation*}
\mathbb{E}\ \omega_{t}^{2n}(\sigma_{j_{.}}\sigma_{k_{.}})=\langle \tilde{q}^{2}_{2n}\rangle_{t}\ , 
\mathbb{E}\ \omega_{t}^{2n}(\tau_{i_{.}}\tau_{l_{.}})=\langle q^{2}_{2n}\rangle_{t}\ , 
\mathbb{E}\ \omega_{t}^{2n}(\tau_{i_{.}}\sigma_{j_{.}})=\langle q_{2n}\tilde{q}_{2n}\rangle_{t}
\end{equation*}
following the last steps of Lemma \ref{lemma} we finally get
\begin{equation}
\label{shift}
\frac{d}{dt}R_M(t)=-\alpha\sum_{n=1}^{\infty}\frac{1}{2n}
\mathbb{E}\tanh^{2n}(\beta J)\langle (q_{2n}-\tilde{q}_{2n})^{2}\rangle_{t}
\end{equation}
Thus
\begin{equation*}
\label{ }
\frac{d}{dt}R_M(t)\leq 0
\end{equation*}
which implies $R_M(1)\leq R_M(0)$ , i.e.
\begin{equation*}
\label{ }
-\beta f_M\leq G_M 
\end{equation*}
for all $\xi$ and $M$, hence
\begin{equation}
\label{bound}
-\beta f_M\leq \inf_{\xi} 
G_M\ \ \ \Box
\end{equation}

Notice that we could obtain the same bounds using any other 
$\eta$ and $\kappa$ leading to the bound (\ref{bound}). 
Even more is true, we could pre-assign the values
$q_{n}$ and forget that they are overlaps of configurations in 
a lattice with $N$ spins, which therefore is not an essential setting. 
Such remark explains the introduction of the following$^{\cite{ass}}$ 
\begin{definition}
A \emph{Random Multi-Overlap Structure} 
$\mathcal{R}$ is a triple 
$(\Sigma, \{q_n\}, \xi)$ where  
\begin{itemize}
  \item $\Sigma$ is a discrete space;
  \item $\xi: \Sigma\rightarrow\mathbb{R}_+$ 
  is a system of random weights; 
  \item $q_n:\Sigma^{n}\rightarrow[0, 1] , n\in\mathbb{N} , |q|\leq 1$
  is a  positive definite \emph{Multi-Overlap Kernel} 
  (equal to 1 only on the diagonal of
  $\Sigma^n$).
\end{itemize}
\end{definition}
Sometimes one considers the closure of $\Sigma$, 
which is not discrete in general.\\
For any RaMOSt one takes a couple of auxiliary 
random variables compatible with the Multi-Overlap Kernel 
and with (\ref{bound}), and the previous theorem 
could be stated$^{\cite{ass}}$ as 
\begin{equation}
\label{rmostbound}
-\beta f_M\leq \inf_{\mathcal{R}} 
G_M
\end{equation}

The generality of the RaMOSt allows to take $\Sigma$
(which is not necessarily
$\{-1, 1\}^{N}$) as the set of indexes of 
the weights $\xi_{\gamma} , \gamma\in\Sigma$ constructed by means 
of Random Probability Cascades of Poisson-Dirichlet processes
(see e.g. ref. \cite{t1}). A well known property of these Cascades 
(see e.g. equation 4.2 in ref. \cite{t1}) gives
place to a chain of expectations of Parisi type 
(see e.g. equation 5.5 in ref. \cite{t1}), that 
coincides with the Parisi Replica Symmetry Breaking theory if one
interpolates according to the iterative approach of ref.
\cite{parisi2}, like in ref. \cite{franz, t1}.
Since we interpolate with different auxiliary Hamiltonians, in our
case (taking identical copies
of $\eta^{\gamma}$ and $\kappa^{\gamma}$
in (\ref{r}) and summing also over $\gamma$) 
the chain of expectations is not equivalent to the Parisi 
Replica Symmetry Breaking in the sense of \cite{parisi2}. This 
point will be deepened elsewhere$^{\cite{dsg}}$. In order to get
the Parisi Replica Symmetry Breaking theory in the traditional sense,
one should take (like in ref. \cite{ass} for non-diluted models) 
a sequence of suitably chosen 
$\eta^{\gamma}$ and $\kappa^{\gamma}$ such that the 
corresponding bounds would be the special 
realization of (\ref{rmostbound})
with the Parisi RaMOSt, like in ref. \cite{ass} for non-diluted models.


\section{Extended Variational Principle}\label{vp}

We can express the free energy of the model 
in the form of the following
\begin{theorem}[Extended Variational Principle]
\begin{equation*}
\label{evp}
-\beta f = \lim_{M\rightarrow\infty}\inf_{\mathcal{R}}G_M
\end{equation*}
\end{theorem}
In order to prove the \emph{Extended Variational Principle}, 
we will find, in Theorem \ref{rb} below, the opposite bound to 
(\ref{rmostbound}), like in ref. \cite{ass}.\\
Notice first that the following Ces{\`a}ro 
limit can be easily computed
\begin{equation}
\label{zeta}
\mathbf{C}\lim_N\frac{1}{M}\mathbb{E}\ln\frac{Z_{N+M}}{Z_N}= -\beta f
\end{equation}
since, 
thanks to the cancellation of the
common terms of the numerator and denominator
\begin{multline}\label{cesaro}
\frac{1}{N}\sum_{K=1}^{N}\frac{1}{M}\mathbb{E}\ln\frac{Z_{K+M}}{Z_K}=\\
\frac{1}{N}\frac{1}{M}(\mathbb{E}\ln Z_{N+M}+\cdots+\mathbb{E}\ln Z_{N+1}
-\mathbb{E}\ln Z_{1}-\cdots-\mathbb{E}\ln Z_{M})
\end{multline}
and the first $M$ terms, with the positive sign, tend to $-\beta f$;
while the others, with the negative sign, vanish in the limit.\\
Theorem \ref{evp} will be proven if we prove the following 
\begin{theorem}[Reversed Bound]
\label{rb}
\begin{equation*}
-\beta f \geq \lim_{M\rightarrow\infty}\inf_{\mathcal{R}}G_M
\end{equation*}
\end{theorem}
\textbf{Proof}\\
If we prove the statement for the restricted RaMOSt space of the 
$\mathcal{R}$'s such that $\xi_\tau$ is$^{\cite{ass}}$ 
the Boltzmann-Gibbs factor  
\begin{equation}
\label{xi}
\xi_\tau=\exp[-\beta(H_N(\tau)+H^{ext}_{N}(\tau))]
\equiv \bar{\xi}_{N}
\end{equation}
then the theorem will hold \textsl{a fortiori}.
Hence, given (\ref{zeta}), it is enough to show
\begin{equation}
\label{revbound}
\mathbf{C}\lim_N\frac{1}{M}\mathbb{E}\ln\frac{Z_{N+M}}{Z_N}\geq
\liminf_N G_M|_{\xi_\tau=\bar{\xi}_{N}}
\end{equation}
We can re-write $G_{M}$ as
$$
\frac{1}{M}\mathbb{E}\ln\left[
\frac{\sum_{\tau, \sigma}\!\exp(-\beta(H_{N}\!+\!\eta))}{Z_{N+M}(\alpha^{\prime})}
\frac{Z_{N+M}(\alpha^{\prime})}{Z_{N+M}(\alpha)}
\frac{Z_{N+M}(\alpha)}{Z_{N}(\alpha)}
\frac{Z_{N}(\alpha)}{\sum_{\tau}\!\exp(-\beta(H_{N}\!+\!\kappa))}\right]
$$
therefore we have four terms. \\
The rest of the proof is very similar to the proof of 
Theorem \ref{lisboa} in the next section.\\
Recall that the sum of Poisson random 
variables is a Poisson random variable
with mean equal to the sum of the means.
So if we take
$$
\alpha^{\prime}=\alpha\frac{(N+M)}{N}
$$
we see that the forth fraction is the 
same as $Z_{N}(\alpha)/Z_{N}(\alpha^{\prime})$.
Furthermore, since 
$\alpha^{\prime}(N+M)-\alpha(N+M)=\alpha^{\prime}N-\alpha N,
\alpha^{\prime}\rightarrow\alpha$,
in the limit the second and forth fractions 
cancel out thanks to Lemma \ref{lemma}
in Appendix \ref{a}. \\
We have seen that the third fraction tends to $-\beta f$. 

Now we proceed with a key step of our approach. 
In the 
denominator of the first fraction we can split the mean
$\alpha^{\prime}(N+M)$ into the sum of three means such that
$H_{N+M}$ splits into the sum of three Hamiltonians 
with the first depending only on
cavity spins $\tau$, the second containing interactions 
between the cavity and the added spins,
the third has the added spins only.
In other words, we are considering the fraction of 
the interactions
(indexed by $\nu$) within the cavity, between the cavity and
added spins, within the added lattice respectively. 
Hence the new three means will be proportional to
$N^{2}, 2NM, M^{2}$ respectively.
More explicitly
\begin{eqnarray*}
\label{ }
&&Z_{N+M}(\alpha^{\prime})=
\sum_{\rho}\exp(\beta\sum_{\nu=1}^{P_{\alpha^{\prime}(N+M)}}
J_{\nu}\rho_{r_{\nu}}\rho_{s_{\nu}})=\\
&&\sum_{\tau , \sigma}\exp\beta(
\sum_{\nu=1}^{P_{\breve{\zeta}}}J_{\nu}\tau_{i_{\nu}}\tau_{l_{\nu}}+
\sum_{\nu=1}^{P_{\tilde{\zeta}}}J_{\nu}\tau_{i_{\nu}}\sigma_{j_{\nu}}+
\sum_{\nu=1}^{P_{\hat{\zeta}}}J_{\nu}\sigma_{j_{\nu}}\sigma_{k_{\nu}})
\end{eqnarray*}
where 
$$
\breve{\zeta}=\alpha^{\prime}\frac{N^{2}}{N+M}\ ,\ 
\tilde{\zeta}=\alpha^{\prime}\frac{2NM}{N+M}\ ,\ 
\hat{\zeta}=\alpha^{\prime}\frac{M^{2}}{N+M}
$$
The third Hamiltonian 
is thus negligible. 
This means that when the cavity is large the
added spins do not interact one another. \\
The choice of $\alpha^{\prime}$ we made guarantees that 
numerator and denominator contain two (up to a 
negligible third in the denominator) identical Hamiltonians with
the same connectivities.
As a consequence, the first fraction in $G_{M}$ 
vanishes in the limit and the theorem 
is proven.  $\Box$


\section{Limiting RaMOSt Invariance}\label{optimal}

As in Appendix \ref{a},
$$
A=\mathbb{E}\ln\cosh(\beta J)-\sum_{n=1}^{\infty}\frac{1}{2n}
\mathbb{E}\tanh^{2n}(\beta J)\langle q_{2n}^{2}\rangle
$$
Then we can state the following
\begin{theorem}\label{lisboa}
In the whole region where the parameters are uniquely
defined, the following Ces{\`a}ro  limit 
is linear in $M$ and $\bar{\alpha}$
\begin{equation}\label{limrost}
\mathbf{C}\lim_{N}\mathbb{E}\ln\Omega_N
\{\sum_{\sigma}\exp[-\beta(\eta(\alpha)+\kappa(\bar{\alpha}))]\}=M(-\beta f 
+\alpha A)+\bar{\alpha}A
\end{equation}
\end{theorem}
\textbf{Proof}\\
The proof is based on the comparison between the limit above and
(\ref{cesaro}), similarly to the method of the previous section.
More precisely, the left hand side of (\ref{limrost}) can be re-written
(without the limit) as
\begin{eqnarray*}
&&\mathbb{E}\ln\Omega_{N}
\Upsilon(0, P_{\alpha N+\bar{\alpha}}, P_{2\alpha M})=\\
&&{}\\
&&\mathbb{E}\ln\frac{\sum_{\tau, \sigma}
\Upsilon(0, P_{\alpha N+\bar{\alpha}}, P_{2\alpha M})}{Z_{N+M}(\alpha^{\prime})}
\frac{Z_{N+M}(\alpha^{\prime})}{Z_{N+M}(\alpha)}
\frac{Z_{N+M}(\alpha)}{Z_{N}(\alpha)}=\\
&&{}\\
&&\mathbb{E}\ln\frac{\sum_{\tau, \sigma}
\Upsilon(0, P_{\alpha N+\bar{\alpha}}, P_{2\alpha M})}
{\sum_{\tau, \sigma}\Upsilon(P_{\alpha^{\prime}\frac{M^{2}}{(N+M)}}, 
P_{\alpha^{\prime}\frac{N^{2}}{(N+M)}}, P_{\alpha^{\prime}\frac{2NM}{(N+M)}})}
\frac{Z_{N+M}(\alpha^{\prime})}{Z_{N+M}(\alpha)}
\frac{Z_{N+M}(\alpha)}{Z_{N}(\alpha)}
\end{eqnarray*}
We know that the third fraction will give $-\beta f M$.
But is is also clear that if we take
\begin{equation*}
\label{ }
\alpha^{\prime}=\frac{(N+M)}{N^{2}}(\alpha N+\bar{\alpha})
\end{equation*}
the three parameters in the numerator of the first fraction tend
to the corresponding ones in the denominator, so that 
the first fraction is immaterial in the limit.
Now notice that
$$
\alpha^{\prime}(N+M)-\alpha (N+M)\rightarrow\alpha M+\bar{\alpha}
$$
and therefore thanks to Lemma \ref{lemma} of Appendix \ref{a}
the contribution of the second fraction is
$$
(\alpha M+\bar{\alpha})A\ \ \ \Box
$$

If we now write
\begin{equation*}
\label{ }
\sum_{\sigma}\exp(-\beta\eta)=
\prod_{j=1}^{M}2\cosh(\beta
\sum_{\nu=1}^{P_{2\alpha}}J_{\nu}^{j}\tau_{i_{\nu}^{j}})
\equiv c_{1}\cdots c_{M}
\end{equation*}
we can formulate (\ref{limrost}) as
\begin{equation*}
\mathbf{C}\lim_{N}\mathbb{E}\ln\Omega_N
\{c_{1}\cdots c_{M}\exp\kappa({\bar{\alpha}})\}=M(-\beta f 
+\alpha A)+\bar{\alpha}A
\end{equation*}
from which it is clear that each cavity field (more precisely
each $c_{j}$)
yields a contribution $(-\beta f +\alpha A)$ in the limit.\\
Notice that in the limiting structure
not only the cavity fields are mutually independent, 
but they are
independent of $\kappa$ as well. 
We have thus obtained the analogy with the 
result of F. Guerra regarding the 
Sherrington-Kirkpatrick model$^{\cite{g2}}$.


\section*{Conclusions}

It is important now to formulate in a complete manner
the Parisi theory for diluted spin glasses$^{\cite{dsg}}$, and
to deepen the analysis of the invariance properties 
of the optimal RaMOSt in order to 
characterize the solution.
We plan to dedicate future work to this program.


\appendix

\section{Connectivity Shift}\label{a}

\begin{lemma}\label{lemma} 
Let $\alpha^{\prime}N=\alpha(N+\Lambda)$, with 
$\Lambda/N\rightarrow 0$ as $N\rightarrow\infty$. Then,
in the whole region where the parameters are uniquely defined
$$
\lim_{N}\mathbb{E}\ln\frac{Z_{N}(\alpha^{\prime})}{Z_{N}(\alpha)}=
\Lambda[\mathbb{E}\ln\cosh(\beta J)-
\sum_{n=1}^{\infty}\frac{\langle q^{2}_{2n}\rangle}{2n}\mathbb{E}\tanh^{2n}(\beta J)]
$$
\end{lemma}
\textbf{Proof}\\
The proof is based on standard convexity arguments.\\
For $t\in[0, 1]$, define
$$
\alpha^{\prime}_{t}=\alpha(1+t\frac{\Lambda}{N})
$$
so that $\alpha^{\prime}_{t}\rightarrow\alpha$ as $N\rightarrow\infty$.\\
We have
$$
A_{t}\equiv\mathbb{E}\ln\frac{Z_{N}(\alpha^{\prime}_{t})}{Z_{N}(\alpha)}=
\mathbb{E}\ln\Omega\exp(-\beta\kappa(\alpha t\Lambda))
$$
Let us compute the $t$-derivative of $A(t)$
\begin{equation*}
\frac{d}{dt}A_{t}=\mathbb{E}\sum_{m=0}^{\infty}\frac{d}{dt}\pi_{\alpha t\Lambda}(m)
\ln\sum_{\tau}\exp(\beta\sum_{\nu=0}^{m}J_{\nu}\tau_{i_{\nu}}\tau_{l_{\nu}})
\end{equation*}
Using the following 
elementary property 
\begin{equation*}\label{poisson}
\frac{d}{dt}\pi_{t\zeta}(m)=\zeta(\pi_{t\zeta}(m-1)-\pi_{t\zeta}(m))
\end{equation*}
we get
\begin{eqnarray*}
&\frac{d}{dt}A_{t}&=
\alpha \Lambda\mathbb{E}\sum_{m=0}^{\infty}[\pi_{\alpha t\Lambda}(m-1)
-\pi_{\alpha t\Lambda}(m)]
\ln\sum_{\tau}\exp(\beta\sum_{\nu=1}^{m}J_{\nu}\tau_{i_{\nu}}\tau_{l_{\nu}})\\
&{}&=\alpha \Lambda\mathbb{E}\ln\sum_{\tau}
\exp(\beta J\tau_{i_{\nu}}\tau_{l_{\nu}})
\exp(\beta\sum_{\nu=1}^{P_{\alpha t \Lambda}}J_{\nu}\tau_{i_{\nu}}\tau_{l_{\nu}})\\
&{}&\hspace{4cm}-\alpha \Lambda\mathbb{E}\ln\sum_{\tau}
\exp(\beta\sum_{\nu=1}^{P_{\alpha t \Lambda}}J_{\nu}\tau_{i_{\nu}}\tau_{l_{\nu}})\\
&{}&=\alpha \Lambda\mathbb{E}\ln\Omega_{t}
\exp(\beta J\tau_{i_{\nu}}\tau_{l_{\nu}})
\end{eqnarray*}
where we included the $t$-dependent weights
in the average $\Omega_{t}$. 
Now use the following identity
$$
\exp(\beta J\tau_{i}\tau_{l})
=\cosh(\beta J)+\tau_{i}\tau_{l}\sinh(\beta J)
$$
to get
\begin{equation*}
\label{ }
\frac{d}{dt}A_{t}=\alpha \Lambda\mathbb{E}\ln\Omega_{t}
[\cosh(\beta J)(1+\tanh(\beta J)\tau_{i_{\nu}}\tau_{l_{\nu}})]
\end{equation*}
We have already observed that 
\begin{equation*}
\label{ }
\mathbb{E}\ \omega_{t}^{2n}(\tau_{i_{.}}\tau_{l_{.}})=\langle q^{2}_{2n}\rangle_{t}
\end{equation*}
so if we now expand the logarithm in power series, we see that 
the result does not depend on $t$ in the limit of large $N$, because 
$\alpha^{\prime}_{t}\rightarrow\alpha$ and hence
$\Omega_{t}\rightarrow\Omega $. Therefore integrating back 
against $t$ from 0 to 1 is the same as multiplying by 1.
Thanks to the symmetric distribution of $J$ we get the result,
where the odd powers are missing. $\Box$


\section{Thermodynamic Limit}\label{tl}

Our new interpolation method should allow one to prove the 
existence of the thermodynamic limit (already proven
in ref. \cite{franz}), for otherwise it would be quite weak. 
As a matter of fact, this theorem turns out to be
elementary with our interpolation.

It is well known that a sufficient condition for the existence of the 
thermodynamic limit is the sub-additivity of the free energy.
In our context we want to measure the changes caused by 
adding the $M$ additional spins to the cavity of size $N$ and make sure 
that
$$
\mathbb{E}\ln Z_{N+M}\geq\mathbb{E}\ln Z_{N}+\mathbb{E}\ln Z_{M}
$$
The natural start is therefore considering the following interpolation
\begin{equation*}
\label{ }
\mathbb{E}\ln\sum_{s}\exp[-\beta(H_{N+M}(t \alpha)+
H_{N}((1-t)\alpha)+H_{M}((1-t)\alpha))]\equiv \Phi_{t}
\end{equation*}
We clearly proceed by splitting $H_{N+M}$ in the usual way, and we get this 
time a total of five Hamiltonians. 
In the $t$-derivative the parts in $\cosh(\beta J)$ cancel out,
since the sum of the five coefficients of $\alpha$ is zero. 
The remaining terms are 
\begin{equation*}
\label{ }
\frac{d}{dt}\Phi_{t}=-\alpha\sum_{n=1}^{\infty}
\frac{\mathbb{E}\tanh^{2n}(\beta J)}{2n}B_{n, t}
\end{equation*}
where
\begin{equation*}
\label{ }
B_{n, t}=
\{\frac{1}{N+M}[\langle (q_{2n}N)^{2}+2q_{2n}N
\tilde{q}_{2n}M+(\tilde{q}_{2n}M)^{2}\rangle_{t}]
-\langle q^{2}_{2n}N+\tilde{q}^{2}_{2n}M\rangle_{t}\}
\end{equation*}
Now it is trivial to show that
$$
B_{n, t}=-\frac{1}{N+M}NM\langle(q_{2n}-
\tilde{q}_{2n})^{2}\rangle_{t}\leq 0 \ \ \Box
$$


\section{Sum Rules and Trial Functions}\label{srtf}

Consider the function $\Phi_t$ defined in the
previous section, then
the fundamental theorem of calculus
implies
$$\mathbb{E}\ln Z_{N+M}-
\mathbb{E}\ln Z_{N}=\mathbb{E}\ln Z_{M}+\int_0^1
\frac{d\Phi_t}{dt}dt
$$ 
Dividing by $M$ and taking the Ces{\`a}ro limit 
as $N$ goes to infinity yields
the following sum rule
$$
-\beta f(\beta, \alpha)=
-\beta f_M(\beta, \alpha)-\alpha
\sum_{n=1}^{\infty}\frac{\mathbb{E}\tanh^{2n}(\beta J)}{2n}
\mathbf{C}\lim_N\int_0^1\langle(q_{2n}-
\tilde{q}_{2n})^2\rangle_t dt
$$
which shows that the difference $f-f_M$ is given in terms
of the multi-overlap distance averaged over any optimal RaMOSt.

If we instead use the fundamental theorem of calculus
for $R_M(t)$ defined in section \ref{bounds}, we get
another sum rule
$$
-\beta f_M = G_M(\mathcal{R},\eta, \kappa)
-\alpha\sum_{n=1}^{\infty}\frac{\mathbb{E}\tanh^{2n}(\beta J)}{2n}
\mathbf{C}\lim_N\int_0^1\langle(q_{2n}-
\tilde{q}_{2n})^2\rangle_t dt
$$
which explains, when $M$ goes to infinity, 
the role of $G_M(\mathcal{R},\eta, \kappa)$
as trial function; minimizing it means finding optimal
RaMOSt's (and therefore the free energy) and
the multi-overlap locking (coalescence) described by Aizenman in the 
case of non-diluted models$^{\cite{aiz}}$.


\section*{Acknoledgments}

The author warmly thanks Francesco Guerra 
for a priceless scientific exchange, and 
would also like to thank Fabio Lucio Toninelli for useful
discussions, Edward Nelson and Yakov Sinai for 
support and availability, Michael Aizenman for encouragement 
and useful conversations, the Department of 
Physics at University of Rome
``La Sapienza'' (and in particular 
Giovanni Jona-Lasinio) for hospitality. 


\end{document}